\RequirePackage{lineno}
\documentclass[10pt]{article}
\usepackage{graphicx}
\usepackage[latin1]{inputenc}
\usepackage{xspace}
\usepackage{subfig}
\usepackage{heppennames2}
\usepackage{wasysym}
\usepackage{amssymb}
\usepackage{url}

\def\Title#1{\begin{center} {\Large #1 } \end{center}}
\def\Author#1{\begin{center}{ \sc #1} \end{center}}
\def\Address#1{\begin{center}{ \it #1} \end{center}}

\newcommand\pubblock{\rightline{\begin{tabular}{l} Proceedings of the Second Annual LHCP\\ \pubnumber\\
         \pubdate  \end{tabular}}}

\newenvironment{Abstract}{\begin{quotation} \begin{center} 
             \large ABSTRACT \end{center}\bigskip 
      \begin{center}\begin{large}}{\end{large}\end{center} \end{quotation}}

\newenvironment{Presented}{\begin{quotation} \begin{center} 
             PRESENTED AT\end{center}\bigskip 
      \begin{center}\begin{large}}{\end{large}\end{center} \end{quotation}}

\def\beq{\begin{equation}}
\def\eeq#1{\label{#1}\end{equation}}
\def\eeqn{\end{equation}}

\def\beqa{\begin{eqnarray}}
\def\eeqa#1{\label{#1}\end{eqnarray}}
\def\eeqan{\end{eqnarray}}

\let\bar=\overbar

\def\Dslash{\not{\hbox{\kern-4pt $D$}}}
\def\dslash{\not{\hbox{\kern-2pt $\del$}}}

\def\Pl{{\mbox{\scriptsize Pl}}}

\def\msb{{\bar{\ssstyle M \kern -1pt S}}}

\newcommand {\pTtrack}   {\ensuremath{p_{\mathrm{T}}^{\mathrm{track}}}\xspace}
\newcommand {\pTjetCharged}     {\ensuremath{p_{\mathrm{T, jet}}^{\mathrm{ch}}}\xspace}
\newcommand {\pT}        {\ensuremath{p_{\mathrm{T}}}\xspace}
\newcommand {\zCh}        {\ensuremath{z^{\mathrm{ch}}}\xspace}
\newcommand {\dEdx}      {d\textit{E}/d\textit{x}\xspace}
\newcommand {\GeVc}      {\ensuremath{\mathrm{GeV}/c}\xspace}
\newcommand {\kT}      {\ensuremath{k_{\mathrm{T}}}\xspace}
\newcommand {\lnull}      {\ensuremath{\tilde{l}}\xspace}
\newcommand {\li}      {\ensuremath{l_{\mathrm{i}}}\xspace}
\newcommand {\lreg}    {\ensuremath{l^{\mathrm{reg}}}\xspace}
\newcommand {\fki}    {\ensuremath{f_{ki}}\xspace}
\newcommand {\lregki}    {\ensuremath{l^{\mathrm{reg}}_{ki}}\xspace}

\usepackage{color}

 \newcommand\pubnumber{ }

\newcommand\pubdate{\today}

\def\affiliation{
On behalf of the ALICE Collaboration, \\
Physikalisches Institut, University of Tübingen, Germany }

\def\support{\footnote{{\it Email address}: \url{Benjamin-Andreas.Hess@Uni-Tuebingen.de}.\\ Work supported by BMBF project 05P12VTCAA.}}

\begin{document}

\large
\begin{titlepage}
\pubblock

\vfill
\Title{ Measurement of hadron composition in charged jets from pp collisions with the ALICE experiment }
\vfill

\Author{ Benjamin Andreas Hess \support }
\Address{\affiliation}
\vfill
\begin{Abstract}

We report on the first measurement of the charged hadron composition in charged jets from pp~collisions. The ALICE detector at the LHC was used to study charged pion, kaon and (anti-)proton production in jets. The results were extracted from $2 \times 10^8$ minimum bias events at a centre-of-mass energy of $\sqrt{s} = 7\,$TeV.

We present the transverse momentum (\pT) spectra and reduced momentum spectra ($\zCh \equiv \pTtrack / \pTjetCharged$) of $\pi$, K and p in charged jets with $\pTjetCharged$ between 5 and 20\,\GeVc. The measurements are compared to Monte Carlo calculations.

\end{Abstract}
\vfill

\begin{Presented}
The Second Annual Conference\\
 on Large Hadron Collider Physics \\
Columbia University, New York, U.S.A \\ 
June 2-7, 2014
\end{Presented}
\vfill
\end{titlepage}
\def\thefootnote{\fnsymbol{footnote}}
\setcounter{footnote}{0}
%

\normalsize 


\section{Introduction}

Jets are phenomenological objects constructed to represent partons originating from hard scattering processes. The present knowledge about parton fragmentation into identified hadrons is mainly constrained by jet fragmentation measurements at $\mathrm{e}^{+}\mathrm{e}^{-}$~colliders \cite{Lee13, Lei13}. In addition, neutral jet fragments have been measured by the CDF Collaboration \cite{Aal13}.
The \mbox{ALICE} experiment \cite{Aam08} at the LHC has powerful particle identification (PID) capabilities allowing for the measurement of identified charged hadron spectra in jets from pp~collisions for the first time.
The charged hadron composition is extracted with sophisticated PID techniques using the specific energy loss (\dEdx) of tracks in the Time Projection Chamber (TPC) \cite{Alm10}. The TPC is the main tracking and PID device of ALICE and has a \dEdx resolution of about 5\% for pp~collisions. The measurement presented here is based on the ALICE analysis strategy for the measurement of inclusive charged particle production in charged jets \cite{Bus13}, but with particle identification using the TPC.

\section{Analysis technique}

The analysis is carried out on a sample of $2 \times 10^8$ minimum bias (MB) events from pp~collisions at a centre-of-mass energy of $\sqrt{s} = 7\,$TeV recorded with the ALICE detector in 2010. It is based on charged tracks with transverse momentum $\pT > 0.15\,\GeVc$ and within the pseudo-rapidity range $|\eta^{\mathrm{track}}| < 0.9$, which are reconstructed with the \mbox{ALICE} Inner Tracking System (ITS) and the TPC. The tracks are clustered to jets using the anti-\kT algorithm of the FastJet \cite{Cac86} package with resolution parameter $R = 0.4$. Only jets with $|\eta^{\mathrm{jet}}| < 0.5$, that are fully contained in the \mbox{ALICE} central barrel acceptance, are included in the analysis.

Based on the TPC \dEdx, the raw $\pi$, K and p yields differential in \pTtrack or $\zCh \equiv \pTtrack / \pTjetCharged$ are extracted. This extraction is done separately for $\pTjetCharged = 5-10$, $10-15$ and $15-20\,\GeVc$.

Finally, the raw yields are corrected for detector efficiency, acceptance, (jet) \pT resolution, secondary particle and muon contamination. The latter correction is required because the TPC \dEdx resolution does not allow for the separation of pions and muons, i.e. the muon yield is attributed to the pion yield. The correction procedure is based on that for the inclusive charged particle measurement described in \cite{Bus13} and has been extended to take into account particle type dependent effects. The following discussion will focus on the PID in jets.

The raw $\pi$, K and p differential yields are extracted with the TPC Coherent Fit \cite{Lu13}. The TPC Coherent Fit is a 2-dimensional fitting procedure that analyses the TPC \dEdx distribution as a function of particle momentum ($p$). It is able to extract raw particle yields differentially in \pT from 0.15\,\GeVc to above 20\,\GeVc with an accuracy of better than 10\% on average for the ALICE pp~collision data. The TPC Coherent Fit is based on the observation that the mean and width of the \dEdx signal as well as the particle fractions of each species are continuous as a function of particle momentum. Using continuous \dEdx models and allowing only for statistical fluctuations of the particle fractions in neighbouring bins, it simultaneously extracts the particle yields and the \dEdx model parameters in a single optimisation procedure.

Technically, the TPC Coherent Fit maximises a log-likelihood function $l$,

\begin{equation}
	l \equiv \lnull + \lreg,
\label{eq:ltot}
\end{equation}
where the additional term \lreg is a regularisation term for the particle yields and will be discussed below. If \fki denotes the particle fraction of species $k$ in momentum bin $i$ and $s_k(p_i;\vec{\Theta})$ is the \dEdx model for species $k$ at momentum $p_i$ with {\it a priori} unknown parameters $\vec{\Theta}$, the log-likelihood term \lnull symbolically reads

\begin{equation}
	\lnull \equiv \sum_i \li \left( \sum_k \fki s_k(p_i;\vec{\Theta}) \right),
\end{equation}
where \li is the log-likelihood function for momentum bin $i$\footnote{The parameter is the modelled \dEdx distribution in that momentum bin. The likelihood that the model describes the measured distribution is evaluated with Poissonian statistics.}. The sum $\sum_i$ adds up the contributions from different momentum bins, whereas the sum $\sum_k$ represents the superposition of the \dEdx distributions from different particle species. The functional form of the particle fraction \fki is not known, but a continuity condition on the particle momentum can be imposed that only allows for statistical deviations of the particle fraction from the interpolated value from the neighbouring momentum bins. This is achieved by adding the regularisation term

\begin{equation}
	\lreg \equiv \sum_{k,i} \lregki(\fki)
\end{equation}
to \lnull in Eq.~\ref{eq:ltot}, where the regularisation strength contributes equally for each particle species and momentum bin\footnote{The regularisation term \lregki is derived from a Gaussian likelihood and, thus, has a proper statistical interpretation.}.

With the \dEdx distribution as a function of momentum as input, the TPC Coherent Fit maximises the log-likelihood in Eq.~\ref{eq:ltot} yielding both the \dEdx model parameters $\vec{\Theta}$ and the particle fractions \fki. The single optimisation procedure is driven {\it coherently} by the full-range constraint of the \dEdx models and the constraint on the particle fractions.

Utilising the \dEdx for PID at high particle momentum ($p > 4\,\GeVc$) is challenging because the extracted particle yields are highly sensitive to the mean \dEdx. Since the K-p separation at such momenta is about 5\% in ALICE pp data, a 1\permil~mean \dEdx bias is estimated to cause a 2\% bias of the particle fractions. Hence, extensive studies of the systematic uncertainties of the TPC Coherent Fit have been performed. These include \dEdx model uncertainties, robustness against changes of the \dEdx quality, a possible particle type dependence of the \dEdx (which is found to be negligible) and \pTjetCharged dependence of the \dEdx. The latter is caused by an increased local track density in jets and automatically taken into account by the fit, since the \dEdx model and fractions are fitted in each \pTjetCharged bin separately.

In addition to these studies, the TPC Coherent Fit was applied to Monte Carlo (MC) samples (PYTHIA \cite{Sjo06} tune Perugia0 \cite{Ska10}) with full ALICE detector simulation (with GEANT3 \cite{Bru93}) and reconstruction in the same way as done for data. Within uncertainties (typically smaller than 10\%), the results of this procedure reproduce the MC truth at the detector level.

Furthermore, the results of the TPC Coherent Fit have been cross-checked by an independent method---the TPC Multi-Template Fit. The TPC Multi-Template Fit extracts in great detail the TPC \dEdx response up to intermediate momenta ($p \lesssim 8\,\GeVc$) from pure MB data samples selected via TPC \dEdx, Time-Of-Flight (TOF) and track topology (products of \PKzS, \PGL and \PAGL decays and of \PGg conversions). The \dEdx response at high momenta is determined from model fits to these clean samples. With this response, templates for the \dEdx distributions of each species are generated for the considered input data sample. Finally, the particle fractions are estimated by minimising the difference between the measured \dEdx distribution and the sum of templates weighted with the particle fractions. The template generation and fitting is performed directly in \pT or $\zCh$ bins, the only free parameters being the particle fractions in each bin. As for the TPC Coherent Fit, the TPC Multi-Template fit uses a log-likelihood maximisation with a regularisation term for the particle fractions.

\begin{figure}[htb]
	\centering
	\includegraphics[height=2.5in]{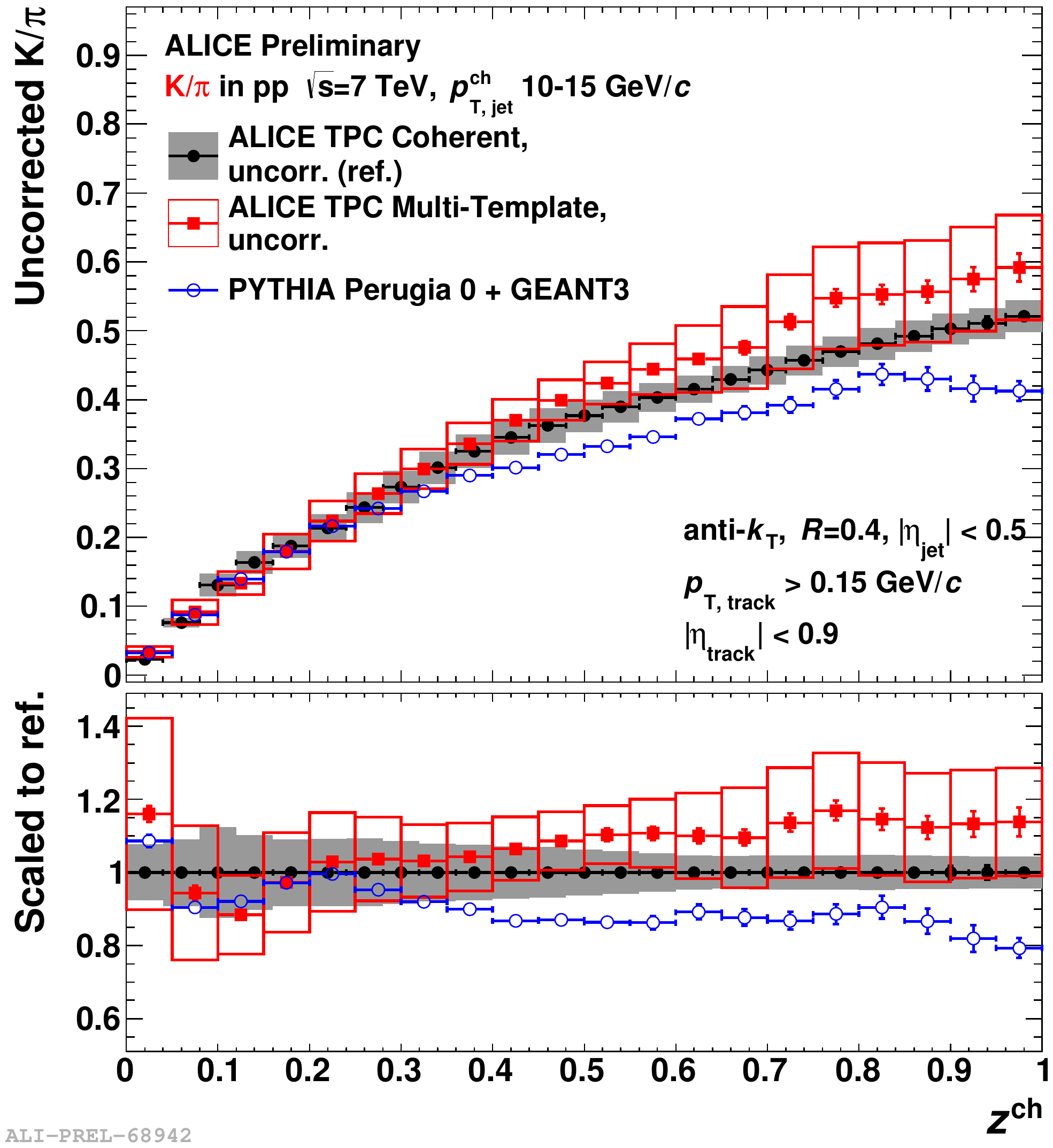}
	
	\caption{Uncorrected K/$\pi$ ratio as a function of \zCh for $\pTjetCharged = 10-15\,\GeVc$ for data and MC (PYTHIA Perugia0). The uncorrected results for data of the TPC Coherent Fit (full points) are compared to those of the TPC Multi-Template Fit (rectangles) and the detector level MC truth (open points). The error bars represent the statistical uncertainties and the error boxes indicate the systematic uncertainties.}
	\label{fig:figure1}
\end{figure}

The two methods have different sources of systematic uncertainties. As shown in Fig.~\ref{fig:figure1}, the results of both methods agree within uncertainties. It can also be seen that PYTHIA Perugia0 MC deviates from data beyond the systematic error.

%

\section{Results}

\begin{figure}[!htb]
	\centering
	\includegraphics[width=\textwidth]{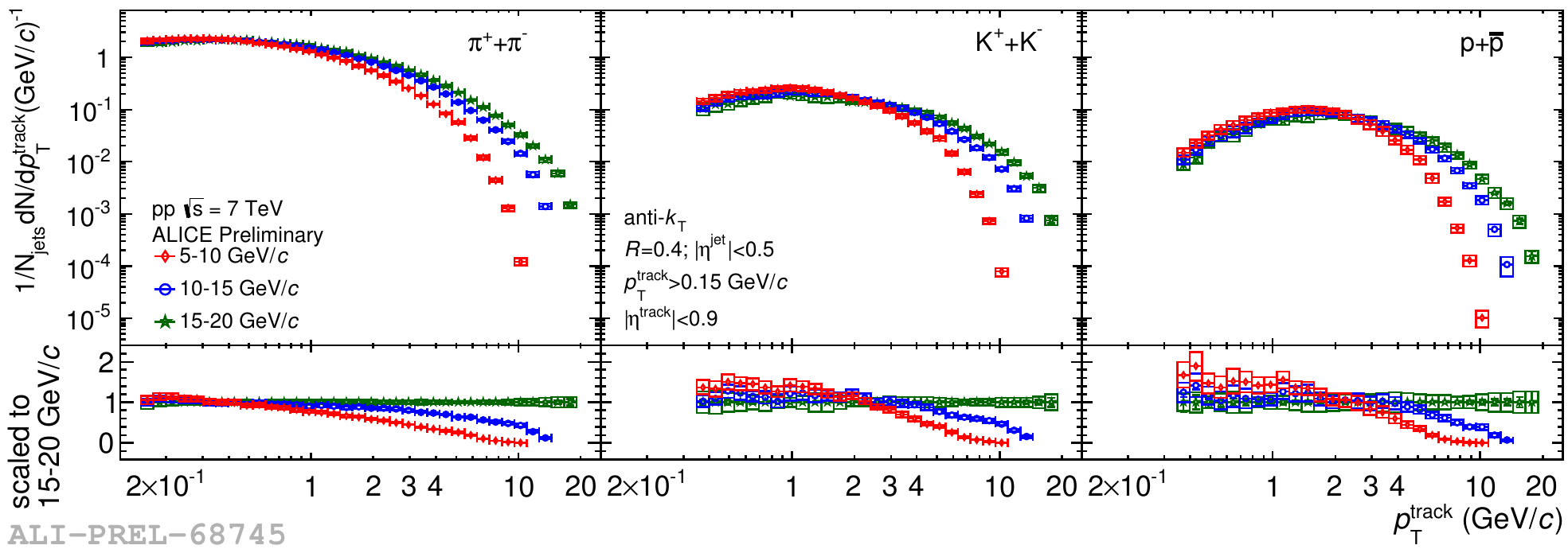}
	
	\caption{Corrected \pT spectra of $\pi$ ({\it left}), K ({\it middle}) and p ({\it right}) in charged jets from pp~collisions at $\sqrt{s} = 7\,$TeV. The spectra for $\pTjetCharged =$ $5-10$ (diamonds), $10-15$ (circles) and $15-20$\,\GeVc (stars) are shown.}
	\label{fig:figure2}
\end{figure}

The fully corrected \pT differential yields per jet of $\pi$, K and p in charged jets are shown in Fig.~\ref{fig:figure2}. The spectra span $3-4$ orders of magnitude and become harder with increasing \pTjetCharged. A clear \pTjetCharged ordering of the spectra is observed; the ordering inverts at $\pT = 0.4\,\GeVc$ for $\pi$ and at $2\,\GeVc$ for K and p.

\begin{figure}[!htb]
	\centering
	\includegraphics[height=2.5in]{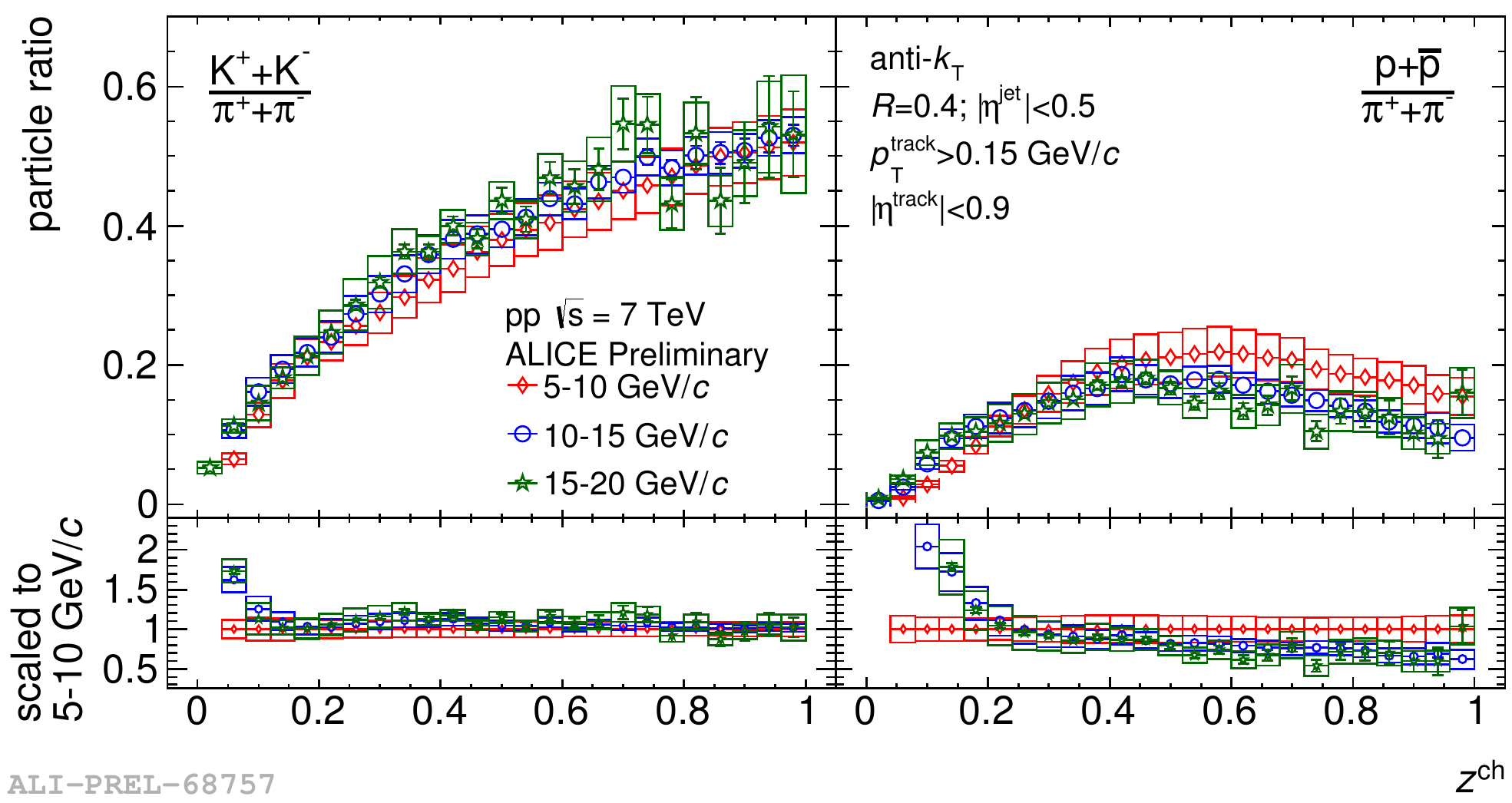}
	
	\caption{Corrected K/$\pi$ ({\it left}) and p/$\pi$ ({\it right}) ratios as a function of \zCh in charged jets from pp~collisions at $\sqrt{s} = 7\,$TeV. The ratios for $\pTjetCharged =$ $5-10$ (diamonds), $10-15$ (circles) and $15-20$\,\GeVc (stars) are shown.}
	\label{fig:figure3}
\end{figure}

In Fig.~\ref{fig:figure3}, the fully corrected K/$\pi$ and p/$\pi$ ratios in charged jets are shown as a function of \zCh. The K/$\pi$ ratio exhibits a monotonic increase with \zCh, reaching $0.5-0.6$, indicating that the strangeness fraction in jets rises with \zCh. For the p/$\pi$ ratio, a maximum of $0.15-0.2$ is reached at $\zCh = 0.5-0.6$ followed by a decrease as \zCh approaches unity. This indicates that leading baryons in charged jets are suppressed. Comparing the ratios for the three different \pTjetCharged bins, a scaling is observed for $\zCh > 0.2$ in all \pTjetCharged bins for the K/$\pi$ ratio and for $\pTjetCharged =$ $10-15$ and $15-20\,\GeVc$ in case of p/$\pi$.

\begin{figure}[!htb]
	\centering
	\includegraphics[width=\textwidth]{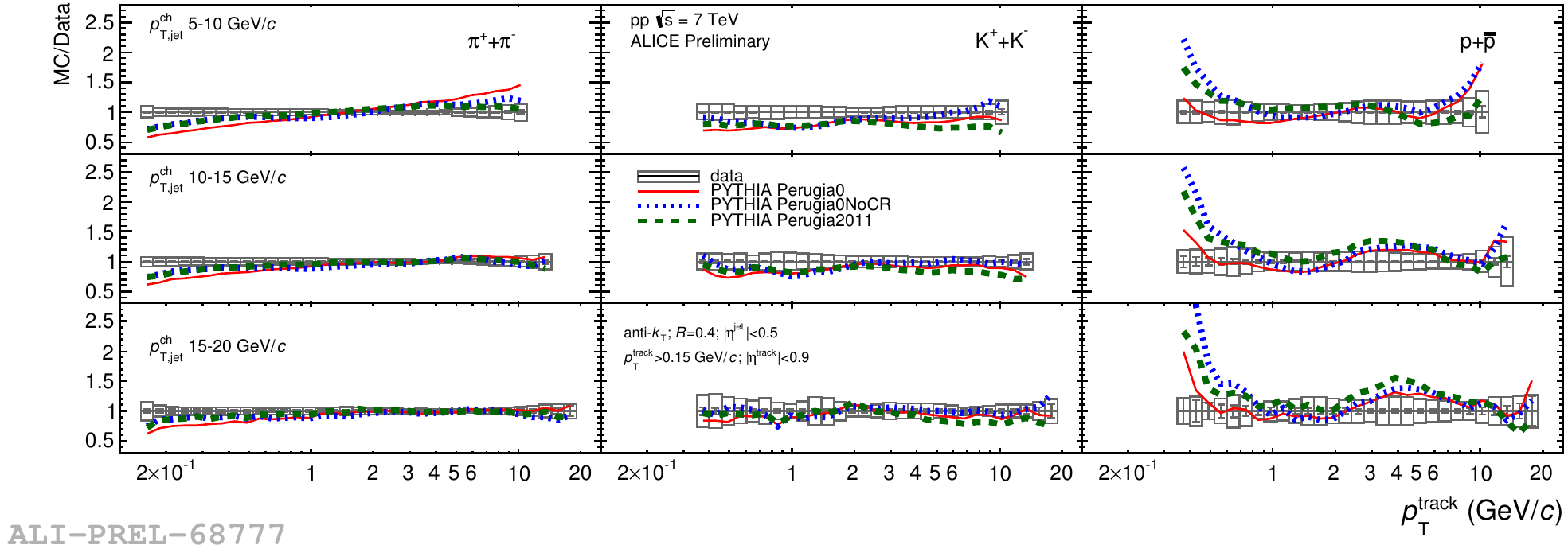}
	
	\caption{MC/data ratios of the $\pi$ ({\it left}), K ({\it middle}) and p ({\it right}) \pT spectra in charged jets from pp~collisions at $\sqrt{s} = 7\,$TeV.}
	\label{fig:figure4}
\end{figure}

In Fig.~\ref{fig:figure4}, the spectra shown in Fig.~\ref{fig:figure2} are compared to the PYTHIA~\cite{Sjo06} tunes Perugia0, Perugia0NoCR ({\it noCR} stands for {\it no colour reconnection}) and Perugia2011 \cite{Ska10} for $\pi$, K and p. The best agreement is observed at high \pTjetCharged and high particle \pT. For low particle \pT, all considered PYTHIA tunes undershoot (overshoot) the pions (protons). Typically, all three tunes describe the data within 30\% except for protons with $\pT < 0.5\,\GeVc$, where the deviation goes beyond 100\%, and with \pT close to the upper bound of the \pTjetCharged bin, where the discrepancy is around 50\% for some tunes. The maximum of the proton spectra around $\pT = 2\,\GeVc$ (cf. Fig.~\ref{fig:figure2}) is reproduced very well by all PYTHIA tunes, but they fail to describe the width and the high-\pT slope. Of the tunes considered, Perugia0NoCR gives the best description of the K spectra in all \pTjetCharged bins.

\section{Conclusions}

We presented the first measurement of identified jet fragmentation of charged hadrons at hadron colliders from ALICE. The particle yields and ratios as functions of \pT and \zCh of primary hadrons ($\pi$, K, p) in charged jets from pp~collisions at $\sqrt{s} = 7\,$TeV with $\pTjetCharged = 5-20\,\GeVc$ are extracted using advanced PID techniques. We observe that the \pTjetCharged scaling of the \zCh spectra disappears at lowest \pTjetCharged. Furthermore, our measurements show an increase of the strangeness fraction with \zCh and a suppression of leading baryons at high \zCh. PYTHIA simulations reproduce the data typically within 30\% accuracy. However, we observe a tension between data and PYTHIA at low \pTjetCharged and for pions and protons at low particle \pT.

\clearpage


\end{document}